\documentclass[aps,preprint,superscriptaddress,showpacs,floatfix]{revtex4}%
\usepackage{amsfonts}
\usepackage{amsmath}
\usepackage{amssymb}
\usepackage{graphicx}
\usepackage{graphicx}%
\begin{document}
\title{Transformation of the frequency-modulated continuous-wave field into a train
of short pulses by resonant filters}
\author{R. N. Shakhmuratov}
\affiliation{Kazan Physical-Technical Institute, Russian Academy of Sciences, 10/7 Sibirsky
Trakt, Kazan 420029 Russia}
\affiliation{Kazan Federal University, 18 Kremlyovskaya Street, Kazan 420008 Russia}
\date{{ \today}}

\begin{abstract}
The resonant filtering method transforming frequency modulated radiation field
into a train of short pulses is proposed to apply in optical domain. Effective
frequency modulation can be achieved by electro-optic modulator or by resonant
frequency modulation of the filter with a narrow absorption line. Due to
frequency modulation narrow-spectrum CW radiation field is seen by the
resonant filter as a comb of equidistant spectral components separated by the
modulation frequency. Tuning narrow-bandwidth filter in resonance with $n$-th
spectral component of the comb transforms the radiation field into bunches of
pulses with $n$ pulses in each bunch. The transformation is explained by the
interference of the coherently scattered resonant component of the field with
the whole comb. Constructive interference results in formation of pulses,
while destructive interference is seen as dark windows between pulses. It is
found that the optimal thickness of the resonant filter is several orders of
magnitude smaller than the necessary thickness of the dispersive filters used
before in optical domain to produce short pulses from the frequency modulated field.

\end{abstract}
\maketitle

\section{Introduction}

Generating pedestal-free optical pulses with high peak power from a low-power
laser is of great interest in optical communication \cite{Nakazawa2000}.
Existing devices generally employ electro-optic amplitude modulators
\cite{Harris2008}, acousto-optic modulators \cite{Warren1,Warren2,Verluise},
frequency chirping followed by dispersive compensators
\cite{Treacy,Grischkowsky1974,Pearson,Nakatsuka,Nikolaus}, and dispersive
modulators \cite{Loy1,Loy2}. Application of the rapid $\pi$-phase-shift
technique for CW radiation field with subsequent filtering by the optically
thick resonant absorber is also capable to produce short pulses in a
controllable way
\cite{Chen2010,Macke,Segard,Shakhmuratov2012,Shakhmuratov2014}. However, this
technique demands very fast phase switch, otherwise the amplitudes of the
pulses reduce appreciably.

Recently, original technique for producing short pulses was reported in
\cite{Vagizov,Shakhmuratov2015}. This technique was experimentally tested with
gamma photons having long coherence length (long duration of a single-photon
wave packet) and the method of splitting of a single photon into pulses
\cite{Vagizov,Shakhmuratov2015} was proposed to create time-bin qubits, whose
concept was introduced before in \cite{Gisin1999,Gisin2002} for optical photons.

The method \cite{Vagizov,Shakhmuratov2015} is based on frequency modulation of
the radiation field, which is also one of the basic elements of the frequency
chirping, implemented in \cite{Grischkowsky1974,Pearson} by electro-optic
modulator. However, in spite of following dispersion compensator, which is
accomplished in \cite{Grischkowsky1974,Pearson} by near resonance absorber
containing alkaline vapor, subsequent absorption (removal) of a particular
spectral component is used \cite{Vagizov,Shakhmuratov2015}. This removal
method is much more flexible compared with the frequency chirping followed by
a dispersive compensators \cite{Grischkowsky1974,Pearson} and allows fine
control of the duration and repetition rate of the pulses.

In this paper we analyze the capabilities of the removal method for
application in optical domain and compare it with dispersion compensator
method. The removal method could be implemented with cold atoms, atomic or
molecular vapors, and organic molecules doped in polymer matrix. They could
serve as filters to remove single frequency component of the comb spectrum
produced by electro-optical modulator from CW radiation field or by modulating
the resonant frequency of the filtering particles. Cold atoms possess almost
naturally broadened Lorentzian absorption lines while atomic vapors at low
pressure demonstrate the Doppler-broadened absorption lines. The effect of the
wings of these lines on the shape of the pulses is discussed. The proposed
method is capable to produce nanosecond or subnanonesocond pulses, which could
be directly resolved in time by modern detectors with no use of
cross-correlation technique, based on Mach-Zehnder interferometer with a delay
line in one of the arms.

I have to mention also theoretical proposals to produce a train of few-cycle femto
and attosecond pulses from vacuum ultraviolet or extreme ultraviolet radiation
fields resonant to an atomic transition in a gas of hydrogenlike atoms,
irradiated by a high-intensity low frequency (LF) laser field far detuned from
all the atomic resonances. \cite{Rad10,Pol11,Antonov13,
Antonov13PRA,Antonov15}. These proposals consider time-dependent perturbation
of the resonantly excited atomic energy level, which oscillates with the
frequency of the LF laser due to Stark effect and ionization.

\section{Basic idea}

CW radiation field $E(t)=E_{0}\exp(-i\omega_{r}t+ikz)$ after passing through
the electro-optic modulator acquires phase modulation%
\begin{equation}
E_{EO}(t)=E(t)e^{im\sin\Omega t}, \label{Eq1}%
\end{equation}
where $\Omega$ and $m$ are the frequency and index of phase modulation.
According to Jacobi-Anger expansion
\begin{equation}
E_{EO}(t)=E(t)%
%TCIMACRO{\dsum \limits_{n=-\infty}^{+\infty}}%
%BeginExpansion
{\displaystyle\sum\limits_{n=-\infty}^{+\infty}}
%EndExpansion
J_{n}(m)e^{im\Omega t}, \label{Eq2}%
\end{equation}
this field is transformed into an equidistant frequency comb with spectral
components $\omega_{n}=\omega_{r}-n\Omega$, where $J_{n}(m)$ is the Bessel
function of the $n$-th order. Fourier transform of the field is
\begin{equation}
E_{EO}(\omega)=E_{0}%
%TCIMACRO{\dsum \limits_{n=-\infty}^{+\infty}}%
%BeginExpansion
{\displaystyle\sum\limits_{n=-\infty}^{+\infty}}
%EndExpansion
J_{n}(m)\delta(\omega-\omega_{n}), \label{Eq3}%
\end{equation}
where $\delta(x)$ is the Dirac delta function. If CW radiation field has
finite spectral width, then $\delta$ function is to be substituted by
$f_{r}(\omega-\omega_{n})$ describing the spectrum of the CW field.

We transmit the frequency comb through the resonant filter with a single
absorption line $F(\omega-\omega_{f})$ centered at frequency $\omega_{f}$. We
select the filter whose absorption linewidth $\Gamma_{f}$ is much smaller than
the distance $\Omega$ between neighboring components of the frequency comb.
Such a filter is capable to remove selectively one of the spectral components
of the comb. Below we don't show explicit spatial dependence of the field
amplitude hiding it into parameters of the filtered field.

By changing the carrier frequency of the CW radiation field we tune the $n$-th
component of the comb $\omega_{n}=\omega_{r}-n\Omega$ close to resonance with
the filter frequency $\omega_{f}$. Then the radiation field at the exit of the
filter is transformed as%
\begin{equation}
E_{fnA}(\omega)=E_{EO}(\omega)+E_{0}J_{n}(m)\left[  T(\Delta_{n})-1\right]
\delta(\omega-\omega_{n}), \label{Eq4}%
\end{equation}
where $\Delta_{n}=\omega_{n}-\omega_{f}$ and $T(\Delta_{n})$ is a transmission
function of the filter, which depends on its absorption coefficient and
physical thickness (see Sec. III). In general, the transmission function
$T(\Delta_{n})$ is a complex function, which takes into account attenuation of
the field amplitude and phase shift due to the frequency dependent refraction
index (dispersion) after passing through the filter of length $L$.

For the optically thick filter at exact resonance $T(0)=T_{0}$ tends to zero.
Indexes $n$ and $A$ in $E_{fnA}(\omega)$ mean that $n$-th component of the
comb is in resonance with the filter and solution is approximated since in Eq.
(\ref{Eq4}) we disregard the interaction of nonresonant components with the
filter. Small contribution from these components due to dispersion will be
taken into account in Sec. III.

Equation (\ref{Eq4}) has a simple physical meaning. It is constructed such
that the first term in square brackets just describes the amplitude of the
attenuated spectral component, which is in resonance and proportional to
$J_{n}(m)T_{0}$. The second term removes from the comb $E_{EO}(\omega)$ the
resonant component to avoid taking it into account twice. Within this
approximation, other spectral components pass through the resonant filter with
no change.

Inverse Fourier transformation of eq. (\ref{Eq4}) results in%
\begin{equation}
E_{fnA}(t)=E_{0}e^{-i\omega_{r}t}\left[  e^{im\sin\Omega t}+(T_{0}%
-1)J_{n}(m)e^{in\Omega t}\right]  . \label{Eq5}%
\end{equation}
To interpret this equation we address the argument given in Feynman lectures
\cite{Feynman}. According to Feynman, the light, transmitted by any sample,
can be considered as a result of the interference of the input wave, as if it
would propagate in vacuum, with the secondary wave radiated by the linear
polarization induced in the medium. Then, following literally this argument,
one can express the output field from the filter $E_{fnA}(t)$ for the input
spectral comb $E_{EO}(t)$ as follows $E_{f}(t)=E_{EO}(t)+E_{sct}(t)$, where
$E_{sct}(t)$ is the field, coherently scattered by the resonant atoms in the
filter, whose amplitude is proportional to $(T_{0}-1)J_{n}(m)$. When filter
becomes opaque for the resonant component $\omega_{n}$ (i.e., $T_{0}%
\rightarrow0$), the amplitude of the coherently scattered field in the forward
direction tends to the amplitude of the resonant component of the input field
and their phases are opposite, i.e. $E_{sct}(t)=-E_{0}J_{n}(m)\exp
(-i\omega_{n}t)$. Destructive interference of the fields results in
attenuation of the $n$-th spectral component of the output field. This concept
has been quantitatively proven for description of optical transients induced
in the optically thick samples by fast phase switch of the incident field,
which reverses destructive interference to constructive one producing short
pulse whose amplitude is two times larger than the amplitude of the incident
radiation field \cite{Shakhmuratov2012}.

Thus, for the frequency comb whose $n$-th spectral component interacts with
the filter, just this component is coherently scattered by the atoms in the
filter. The scattered field interferes with the whole frequency comb at the
exit of the filter. Therefore the output radiation field reveals unusual properties.

The intensity of the field at the exit of the filter $I_{fnA}(t)=\left\vert
E_{fnA}(t)\right\vert ^{2}$ is described by equation%
\begin{equation}
I_{fnA}(t)=I_{0}\left[  1-2S_{n}\cos\psi_{n}(t)+S_{n}^{2}\right]  ,
\label{Eq6}%
\end{equation}
where $I_{0}=\left\vert E_{0}\right\vert ^{2}$, $S_{n}=(1-T_{0})J_{n}(m)$ and
$\psi_{n}(t)=n\Omega t-m\sin\Omega t$. If the filter is opaque for the
resonant component, the amplitude of the scattered field almost coincides with
the amplitude of the resonant component and $S_{n}\rightarrow J_{n}(m)$. The
phases of these fields are opposite, therefore, if $\cos\psi_{n}(t)$ is
positive we have destructive interference seen as a drop of intensity
$I_{fnA}(t)$. If $\cos\psi_{n}(t)$ is negative, the comb and the scattered
resonant component interfere constructively producing pulses. The interference
becomes pronounced if $J_{n}(m)$ has global maximum. For different spectral
components $n$ this maximum is achieved at different values of the modulation
index $m$. We denote these values as $m_{n}$, which are $m_{1}=1.8$,
$m_{2}=3.1$, $m_{3}=4.2$, ... The phase difference of the comb $E_{EO}(t)$
(whose phase evolves as $m\sin\Omega t$) and the scattered field $E_{sct}(t)$
(whose phase evolves as $n\Omega t$) is $\psi_{n}(t)+\pi$ if $J_{n}(m)>0$. The
evolution of $\psi_{n}(t)$ in time fully describes the formation of the pulses
and the dark windows between them. Time evolution of the phase $\psi_{n}(t)$
and subsequent formation of the pulses for $n=1$, $2$, and $3$, are shown in
Fig. 1 (a, b, and c, respectively). \begin{figure}[ptb]
\resizebox{0.5\textwidth}{!}{\includegraphics{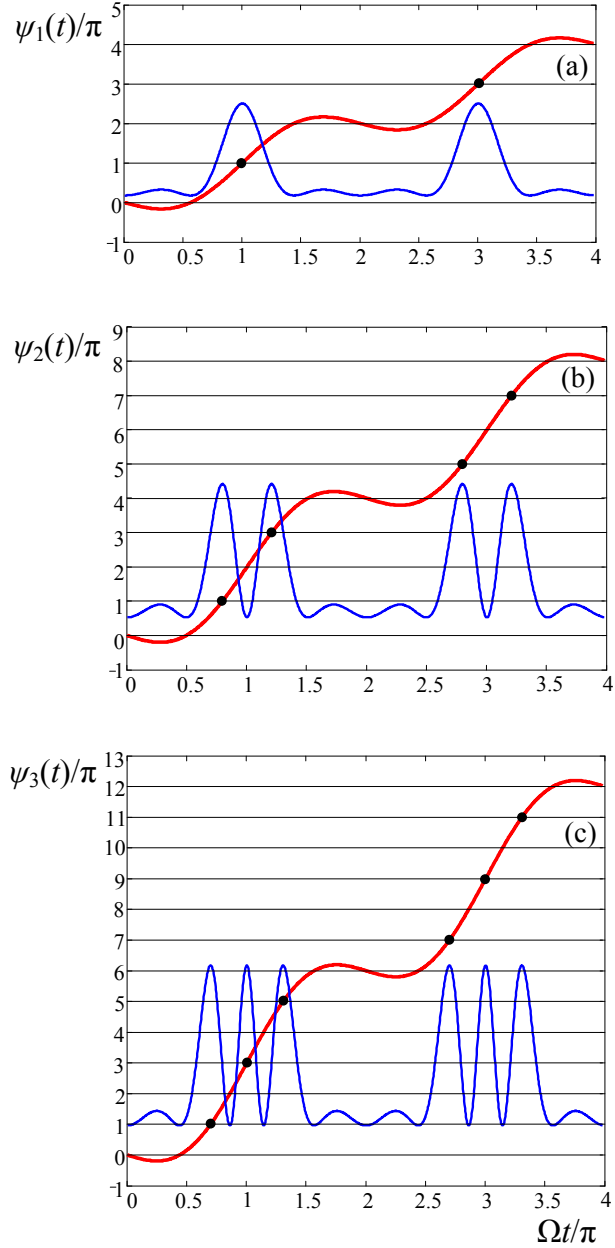}}\caption{(color on
line) Time evolution of the phase difference of the comb and resonantly
scattered field component, $\psi_{n}(t)$, for $n=1$ (a), $n=2$ (b), and $n=3$
(c), thick line in red. $\pi$-pase shift is not shown in the phase difference
for simplicity of notations. Black circles indicate the points when $\psi
_{n}(t)=(2k+1)\pi$, where $k$ is integer. The optimal values of the modulation
index are taken, i.e., $m_{1}=1.8$, $m_{2}=3.1$, and $m_{3}=4.2$ (see the
text). Thin solid line in blue shows the formation of pulses according to Eq.
(\ref{Eq6}). For visualization the intensity of the pulses are scaled to fit a
half of each plate with no reference to the actual level of zero intensity.
Numerical values of the pulse intensities and pedestal level are given in the
text.}%
\label{fig:1}%
\end{figure}

We see that the phase modulated field $E_{EO}(t)$ after passing through the
resonant filter is transformed into bunches of pulses. The number of pulses in
a bunch is equal to the number of filtered component of the frequency comb
$n$. The intensity of the pulses is equal to $[1+S_{n}(t)]^{2}I_{0}$. For
example, the maximum intensities, predicted by Eq. (\ref{Eq6}) for optimal
values of the modulation index $m_{n}$ and opaque filter ($T_{0}=0$), are
$I_{\max}=2.5I_{0}$ for $n=1$, $I_{\max}=2.1I_{0}$ for $n=2$, and $I_{\max
}=2.06I_{0}$ for $n=3$. Thus, the intensity of the pulses exceeds almost two
times the intensity of the radiation field if it would not interact with the
filter. The radiation intensity between bunches is quite small because of
destructive interference, which predicts $I_{\min}=[1-S_{n}(t)]^{2}I_{0}$. For
the same values of the parameters $T_{0}$ and $m_{n}$ this intensity is
$I_{\min}=0.175I_{0}$ for $n=1$, $I_{\min}=0.26I_{0}$ for $n=2$, and $I_{\min
}=0.32I_{0}$ for $n=3$, i.e., almost an order of magnitude smaller compared
with the pulse maxima.

Qualitatively the pulse bunching can be understood from the time evolution
analysis of the phase $\psi_{n}(t)$. Within the period of the phase
oscillation ($T_{EO}=2\pi/\Omega$), induced by electro-optical modulator, one
can distinguish two intervals. Half of the period $T_{EO}/2$ the function
$\psi_{n}(t)$ evolves almost linearly as $(n+m)\Omega t+C$ where $C$ is
constant but different for different bunches (see Fig. 1). Within this half of
the period the relation $\psi_{n}(t)=(2k+1)\pi$, where $k$ is integer, is
satisfied $n$-times (shown by black circles in Fig. 1), producing pulses.
During the other half of the period $T_{EO}/2$ the evolution of phase
$\psi_{n}(t)$ almost stops near the value $2k\pi$, resulting in destructive
interference, seen as dark windows.

\section{Filtering trough cold atoms}

In this section we consider the frequency comb filtering through laser-cooled
atoms with a modest optical depth. As an example we take parameters of $^{85}%
$Rb atoms in a two-dimensional magneto-optical trap, described, for example,
in \cite{Chen2010}. CW radiation field excites $^{85}$Rb D1-line transition
($\lambda=795$ nm). Since the absorption line is almost Lorentzian, the
transmission function in Eq. (\ref{Eq4}) can be described as \cite{Crisp}
\begin{equation}
T(\Delta_{n})=\exp\left(  -\frac{\alpha L\gamma/2}{\gamma-i\Delta_{n}}\right)
, \label{Eq7}%
\end{equation}
where $\alpha$ is the Beer's law absorption coefficient, $L$ is the length of
atomic cloud, and $\gamma$ is a halfwidth of absorption line. If $n$-th
component of the frequency comb is in exact resonance with D1 line, then the
transmission function is $T(0)=\exp(-\alpha L/2)$. Atomic cloud with a length
of few millimeters demonstrates already optical depth $\alpha L=5$.

One can verify the approximate expression for the filtered field $E_{fnA}(t)$
(\ref{Eq5}) comparing it with the exact expression, which could be obtained by
the convolution of $E_{EO}(t)$ field
\begin{equation}
E_{fn}(t)=%
%TCIMACRO{\dint \limits_{-\infty}^{+\infty}}%
%BeginExpansion
{\displaystyle\int\limits_{-\infty}^{+\infty}}
%EndExpansion
E_{EO}(t-\tau)R(\tau)d\tau, \label{Eq8}%
\end{equation}
with the Green's function of the absorber of thickness $L$
\cite{Shakhmuratov2012,Shakhmuratov2014,Crisp}
\begin{equation}
R(t)=\delta(t)-\Theta(t)e^{-(i\omega_{f}+\gamma)t}j_{1}(bt) \label{Eq9}%
\end{equation}
where $\delta(t)$ is the Dirac $\delta$ function, $\Theta(t)$ is the Heaviside
step function, $j_{1}(bt)=\sqrt{b/t}J_{1}\left(  2\sqrt{bt}\right)  $,
$J_{1}(x)$ is the Bessel function of the first order, and $b=\alpha L\gamma
/2$. For the infinitely lasting field, Eq. (\ref{Eq8}) is reduced to
\begin{equation}
E_{fn}(t)=E_{EO}(t)-E(t)%
%TCIMACRO{\dint \limits_{0}^{+\infty}}%
%BeginExpansion
{\displaystyle\int\limits_{0}^{+\infty}}
%EndExpansion
j_{1}(b\tau)e^{(i\Delta-\gamma)\tau+im\sin\Omega(t-\tau)}d\tau. \label{Eq10}%
\end{equation}
where $\Delta=\omega_{r}-\omega_{f}$. Index $n$ in $E_{fn}(t)$ implies that we
consider the case when $n$-th spectral component of the comb is close to
resonance with the filter, i.e., $\Delta=n\Omega+\Delta_{n}$ and $\Delta_{n}$
is close to zero.

Equation (\ref{Eq10}) takes into account the transformation of all spectral
components of the comb including those whose change is infinitely small since
they are far from resonance with the filter. Comparison of the time
dependencies of the approximate intensity $I_{fnA}(t)$, Eq. (\ref{Eq6}), and
exact expression $I_{fn}(t)=\left\vert E_{fn}(t)\right\vert ^{2}$, where
$\Delta=n\Omega$ and $\Delta_{n}=0$, is shown in Fig. 2 for $n=1$, $2$, and
$3$. Optimal values of the modulation index $m=m_{n}$ are adopted in each
case. The following parameters for the cold-atom filter: $\gamma/2\pi=3$ MHz,
$\alpha L=5$, and $b/2\pi=7.5$ MHz, are taken. We choose the modulation
frequency of the field phase equal to $30$ MHz, satisfying well the condition
that $\Omega$ is much larger than the spectral width of absorption line of the
filter $\Gamma_{f}=2\gamma$.

In Fig. 2(a) pulse duration from shoulder to shoulder for $n=1$ is slightly
shorter than half of the period of the phase oscillation $T_{EO}/2=16.7$ ns,
while duration of the dark window is slightly longer than $T_{EO}/2$ [see also
Fig. 1(a)]. The pulsewidth at halfmaximum is close to but slightly shorter
than $T_{EO}/4=8.3$ ns. If we have $n>1$ pulses in a bunch (see Fig. 1 and
Fig. 2), their pulsewidth can be roughly estimated as $T_{EO}/4n$. Therefore,
if, for example, $n=3$, the pulsewidth is already close to $1$ ns.
\begin{figure}[ptb]
\resizebox{0.5\textwidth}{!}{\includegraphics{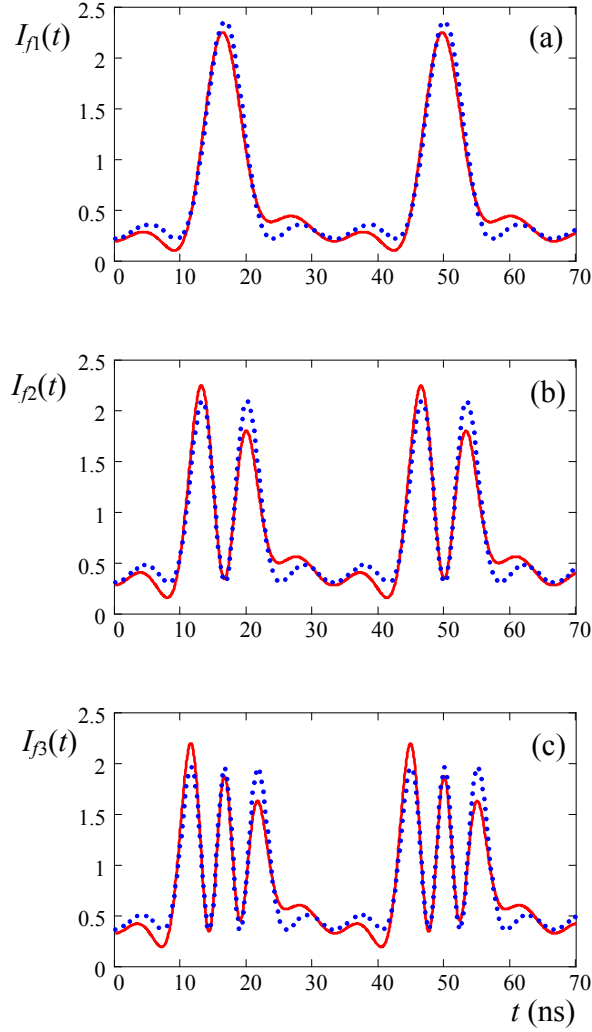}}\caption{(color on
line) Time dependence of the intensity of the filtered radiation field. Solid
line in red corresponds to exact expression (\ref{Eq10}) and dotted line in
blue represents the analytical approximation (\ref{Eq6}). Both are normalized
to the intensity of the incident radiation field $I_{0}$. The modulation
frequency $\Omega/2\pi=30$ MHz is ten times larger than the halfwidth of the
absorption line of the filter $\gamma/2\pi=3$ MHz, the optical depth of the
filter is $\alpha_{0}L=5$. The values of the modulation index and the number
of the spectral component $n$, tuned in resonance ($\Delta_{n}=0$), are the
same as in Fig. 1.}%
\label{fig:2}%
\end{figure}

Small misfit between exact and approximate time dependencies is caused by the
contribution of two nearest neighbors $\omega_{n\pm1}=\omega_{r}-(n\pm
1)\Omega$ of the resonant spectral component $\omega_{n}$. The corrected
expression, which takes them into account, is easily found%
\begin{equation}
E_{fn}(t)\approx E_{fnA}(t)+E_{n+1}(t)+E_{n-1}(t), \label{Eq11}%
\end{equation}
where%
\begin{equation}
E_{n\pm1}(t)=E_{0}e^{-i\omega_{r}t+i(n\pm1)\Omega t}J_{n\pm1}(m_{n})\left(
e^{\frac{-b}{\gamma-i(\Delta_{n}\mp\Omega)}}-1\right)  . \label{Eq12}%
\end{equation}
The field intensity, calculated with this correction, describes almost
excellent the exact time dependence of the filtered field intensity
$I_{fn}(t)$. As it is seen in Fig.2, the contribution of the sidebands
introduces the asymmetry of the pulse intensities within a bunch and
after-wringing, which appears in the beginning of dark windows.
\begin{figure}[ptb]
\resizebox{0.5\textwidth}{!}{\includegraphics{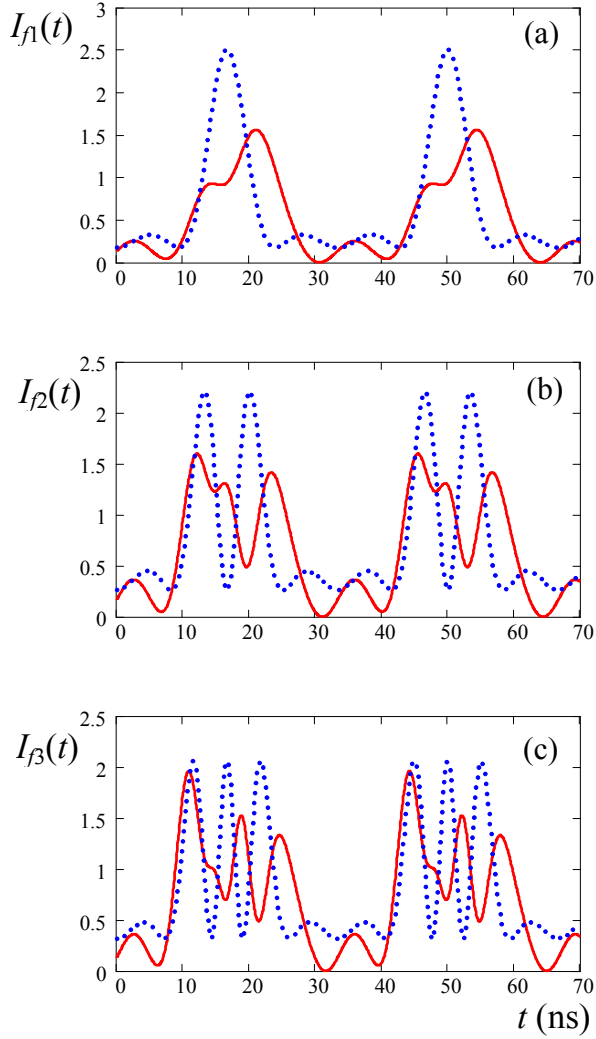}}\caption{Time
dependence of the intensity of the filtered radiation field for atomic cloud
of the length $L=1.5$ cm, which corresponds to the optical depth $\alpha
_{0}L=33$ \cite{Chen2010}. Other parameters and notations are the same as in
Fig. 2}%
\label{fig:3}%
\end{figure}\begin{figure}[ptbptb]
\resizebox{0.5\textwidth}{!}{\includegraphics{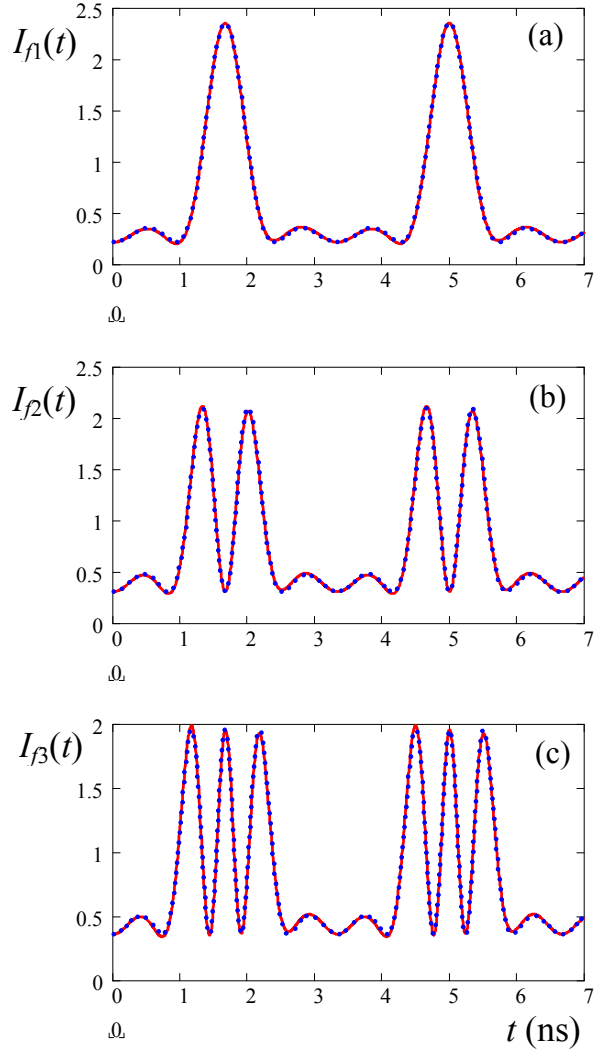}}\caption{Comparison of
the exact time dependence of the intensity of the filtered radiation field
$I_{fn}(t)$ (solid line in red) with the approximate one $I_{fnA}(t)$ (blue
dots) for atomic cloud with optical depth $\alpha_{0}L=5$. Modulation
frequency is $\Omega/2\pi=300$ MHz. Other parameters and notations are the
same as in Fig. 2}%
\label{fig:4}%
\end{figure}\begin{figure}[ptbptbptb]
\resizebox{0.5\textwidth}{!}{\includegraphics{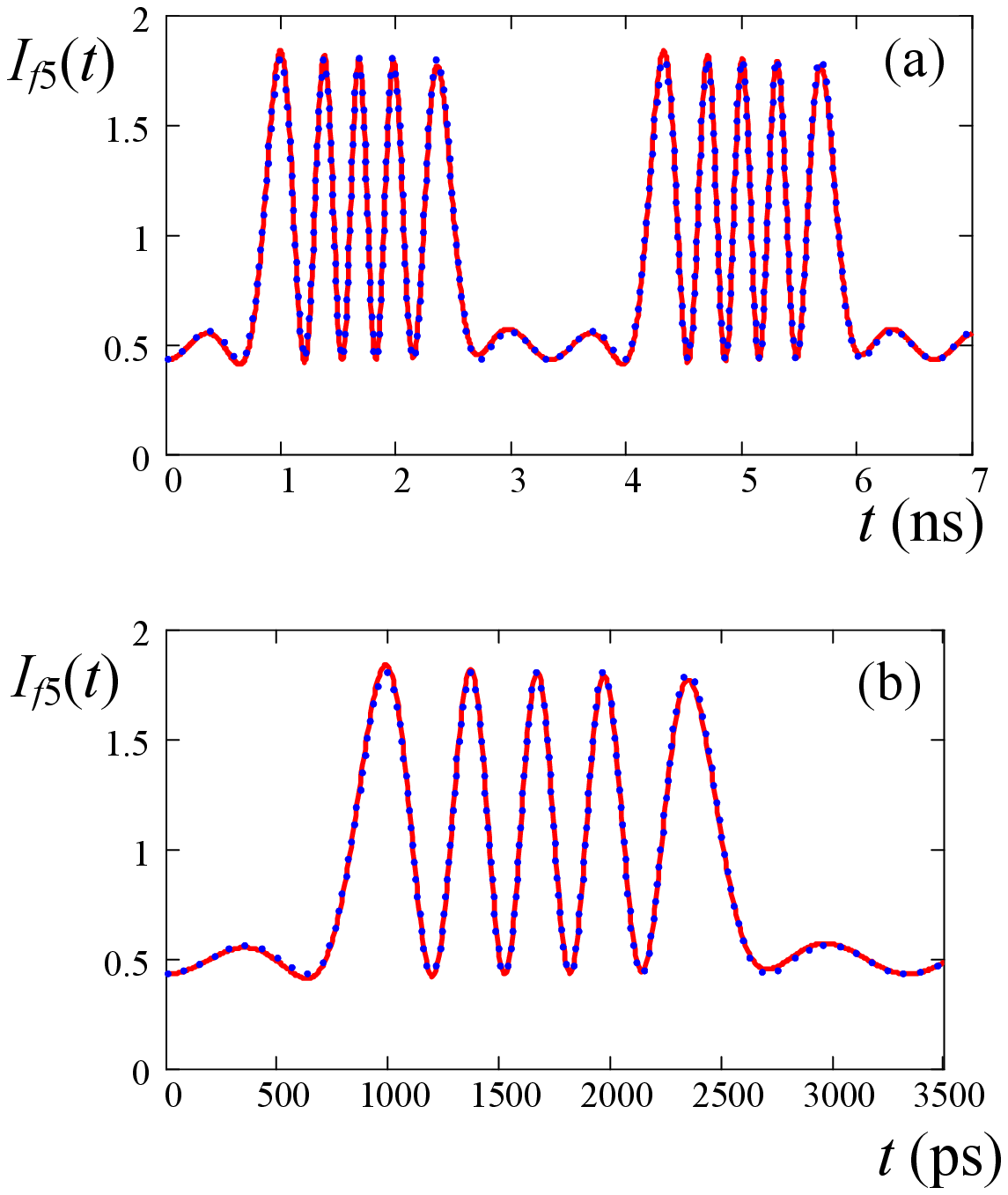}}\caption{(a) Time
evolution of intensity of the filtered radiation field for atomic cloud with
optical depth $\alpha_{0}L=5$. Modulation frequency is $\Omega/2\pi=300$ MHz.
The sideband $n=5$ is tuned in resonance with the filter. The value of the
modulation index $m_{5}=6.4$ is close to $2\pi$. (b) Zoom in the first bunch
of pules, shown in (a). Solid line in red shows $I_{fn}(t)$ and blue dots
correspond to $I_{fnA}(t)$. Both are normalized to $I_{0}$.}%
\label{fig:5}%
\end{figure}

Parameter $b=\alpha L\gamma/2$ plays a crucial role in the radiation filtering
since the transmission function $T(\Delta)=\exp\left[  -b/(\gamma
-i\Delta)\right]  $ essentially broadens if $b$ becomes larger than $\gamma$.
In this case many spectral components of the comb are modified by the filter
and the interference of the comb with the scattered radiation field becomes
messy. Figure 3 shows a comparison of the time dependencies of exact intensity
$I_{fn}(t)$ and approximate one $I_{fnA}(t)$ for atomic cloud with optical
depth $\alpha L=33$, which is achieved in \cite{Chen2010} with cloud length
$L=1.5$ cm. Instead of nice pulses we see their appreciable distortion. This
is not surprising since for this cloud the parameter $b/2\pi=49.5$ MHz is
larger than the frequency comb spacing $\Omega/2\pi=30$ MHz and many sidebands
of the resonant component contribute to the pulse generating. For these values
of the parameters it is necessary to take 8 neighboring components into
account, i.e., four red detuned from resonance and four blue detuned. Then,
the approximate expression similar to Eq. (\ref{Eq11}) but with eight
additional terms $E_{n\pm k}(t)$, where $k=1$, 2, 3, and 4, gives the same
result as exact equation (\ref{Eq10}). Actually the exact equation can be
expressed as a result of filtering of all spectral components of the comb
\begin{equation}
E_{fn}(t)=E_{fnA}(t)+%
%TCIMACRO{\dsum \limits_{k=1}^{\infty}}%
%BeginExpansion
{\displaystyle\sum\limits_{k=1}^{\infty}}
%EndExpansion
\left[  E_{n+k}(t)+E_{n-k}(t)\right]  , \label{Eq13}%
\end{equation}
where%
\begin{equation}
E_{n\pm k}(t)=E_{0}e^{-i\omega_{r}t+i(n\pm k)\Omega t}J_{n\pm k}(m_{n})\left(
e^{\frac{-b}{\gamma-i(\Delta_{n}\mp k\Omega)}}-1\right)  . \label{Eq14}%
\end{equation}

Contrary to the example of the optically thick filter modifying many spectral
components of the comb, we give another example when filtering becomes ideal.
We take moderate optical depth $\alpha L=5$, which corresponds to $b/2\pi=7.5$
MHz, and increase modulation frequency ten times to the value $\Omega
/2\pi=300$ MHz. Then, the contribution of the sidebands becomes negligible and
the intensity of the filtered radiation is well described by approximate
equation (\ref{Eq6}) (see Fig. 4).

It is interesting to note that the phase modulation with frequency 300 MHz is
capable to produce subnanosecond pulses. Tuning, for example, the 5-th
sideband of the frequency comb into resonance with the filter is capable to
produce pulses with duration of 167 ps (see Fig. 5).

Cold $^{85}$Rb atoms are not the only example of the narrow bandwidth filter.
One can use also a cloud of cold potassium ($^{39}$K) atoms generated in a
vapor-cell magneto optic trap whose excitation on the $4S_{1/2}%
(F=1)\leftrightarrow4P_{1/2}(F=2)$ transition (transition wave-length 770 nm)
was studied in \cite{Jeong} for observation of optical precursors.

\section{Filtering trough atomic vapor}

In this section we consider the filtering of the frequency comb through a
vapor of $^{87}$Rb atoms and take the parameters of the experiment
\cite{Lukin} where spectral properties of the electromagnetically induced
transparency were studied. Assume that the fundamental frequency of the comb
is close to the $S_{1/2},F=1\rightarrow P_{1/2},F=2$ transition of the $D_{1}$
line of natural Rb ($\lambda=795$ nm). The atoms are confined in a cell of
length $L=5$ cm. We take two temperatures of Rb vapor (50 and 70$^{\circ}$C),
which correspond to atomic densities $N_{1}=6\times10^{10}$ cm$^{-3}$ and
$N_{2}=6\times10^{11}$ cm$^{-3}$, respectively. Natural linewidth of the Rb
$D_{1}$ line is $\Gamma_{f}/2\pi=\gamma/\pi=5.4$ MHz and Doppler broadening is
$\Delta\omega_{D}/2\pi=500$ MHz. We take the phase modulation frequency
$\Omega/2\pi=10$ GHz, which is 20 times larger than the Doppler width
$\Delta\omega_{D}/2\pi=500$ MHz.

The transmission function for the atomic vapor is%
\begin{equation}
T_{D}(\Delta_{n})=\exp[-\alpha_{1,2}LF_{D}(\Delta_{n})/2], \label{Eq15}%
\end{equation}
where $\alpha_{1,2}=3N_{1,2}\lambda^{2}/2\pi$ is the absorption coefficient of
the naturally broadened line and $F_{D}(\Delta_{n})$ is a Doppler broadened
absorption line, which is
\begin{equation}
F_{D}(\Delta_{n})=\sqrt{\frac{\ln2}{\pi}}\frac{2\gamma}{\Delta\omega_{D}}%
%TCIMACRO{\dint \limits_{-\infty}^{+\infty}}%
%BeginExpansion
{\displaystyle\int\limits_{-\infty}^{+\infty}}
%EndExpansion
\frac{\exp[-\ln2(2x/\Delta\omega_{D})^{2}]}{\gamma-i(\Delta_{n}+x)}dx.
\label{Eq16}%
\end{equation}
We have to emphasize that for both densities of Rb atoms the cell is optically
thick, i.e., $\alpha_{1}L=905$ and $\alpha_{2}L=9053$. It is easy to show
(see, for example, \cite{Shakhmuratov2008}) that at exact resonance
($\Delta_{n}=0$) Eq. (\ref{Eq16}) can be approximated as%
\begin{equation}
F_{D}(0)=\sqrt{\pi\ln2}\frac{2\gamma}{\Delta\omega_{D}}. \label{Eq17}%
\end{equation}
Therefore for $\Delta_{n}=0$ the effective optical depth, $\alpha_{1,2}%
LF_{D}(0)$, is reduced almost hundred times since the Doppler width
$\Delta\omega_{D}$ is two orders of magnitude larger than the natural
linewidth $2\gamma$.

If $\left\vert \Delta_{n}\right\vert >1.8\Delta\omega_{D}$, then the
absorption line reveals Lorenzian wings (see, for example, Fig. 1 in Ref.
\cite{Shakhmuratov2008}), which can be approximated as%
\begin{equation}
F_{D}(\Delta_{n})=\frac{\gamma}{\gamma-i\Delta_{n}}. \label{Eq18}%
\end{equation}
Therefore, the contribution of far wings of the dispersion $\chi^{\prime
}(\Delta_{n\pm k})\sim\operatorname{Im}F_{D}(\Delta_{n\pm k})$ in the
filtering of nonresonant components $\omega_{r}-(n\pm k)\Omega$ of the
frequency comb could be noticeable if $\Delta_{n}=0$. For example, the
contribution of the nearest sidebands ($k=\pm1$) of the resonant component is
proportional to
\begin{equation}
E_{n\pm1}(t)\sim J_{n\pm1}(m_{n})\left(  e^{\mp ib_{1,2}/\Omega}-1\right)  ,
\label{Eq19}%
\end{equation}
where $b_{1,2}=\alpha_{1,2}\gamma/2$. For the atomic density $N_{1}$ the ratio
$b_{1}/\Omega=0.122$ is small since $b_{1}/2\pi=1.2$ GHz and modification of
the sidebands due to filtering does not influence significantly the shape of
the produced pulses. For the atomic density $N_{2}$ the ratio $b_{2}%
/\Omega=1.22$ is large since $b_{2}/2\pi=12$ GHz and the sidebands change
their phases due to filtering. In this case one can expect appreciable
corruption of the produced pulses. To verify these expectations we calculate
the intensity of the filtered comb taking into account the modification of
many sidebands. The number of them can be limited by $\pm k_{\max}$ if the
contribution of the next sidebands $\pm(k_{\max}+1)$ does not change the signal.

Substituting the transmission function $T_{D}(\Delta_{n})$ into equations
(\ref{Eq5}) and (\ref{Eq14}) one obtains the modified Eq. (\ref{Eq13}), which
describes the transformation of the frequency comb after passing through the
atomic-vapor filter. The substitution changes the functions $S_{n}$ and
$E_{n\pm k}(t)$ to%
\begin{equation}
S_{n}=J_{n}(m_{n})[1-T_{D}(0)], \label{Eq20}%
\end{equation}%
\begin{equation}
E_{n\pm k}(t)=E_{0}e^{-i\omega_{r}t+i(n\pm k)\Omega t}J_{n\pm k}(m_{n})\left[
T_{D}(\mp k\Omega)-1\right]  , \label{Eq21}%
\end{equation}
where $\Delta_{n}=0$, which implies that the $n$-th spectral component is in
exact resonance with the filter. \begin{figure}[ptb]
\resizebox{0.5\textwidth}{!}{\includegraphics{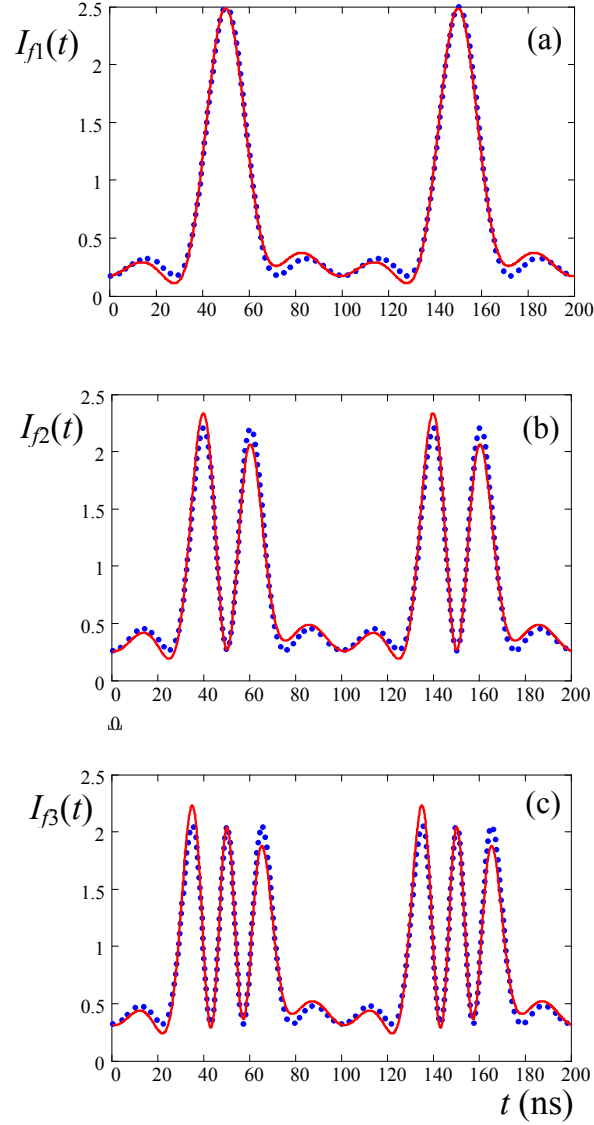}}\caption{Time
dependence of the intensity of the filtered radiation field $I_{fn}(t)$ (solid
line in red) for atomic vapor with optical depth $\alpha_{1}L=905$. Modulation
frequency is $\Omega/2\pi/2\pi=10$ GHz. The approximate time dependence of the
intensity $I_{fnA}(t)$ is shown by blue dots. Other parameters and notations
are the same as in Fig. 2}%
\label{fig:6}%
\end{figure}\begin{figure}[ptbptb]
\resizebox{0.5\textwidth}{!}{\includegraphics{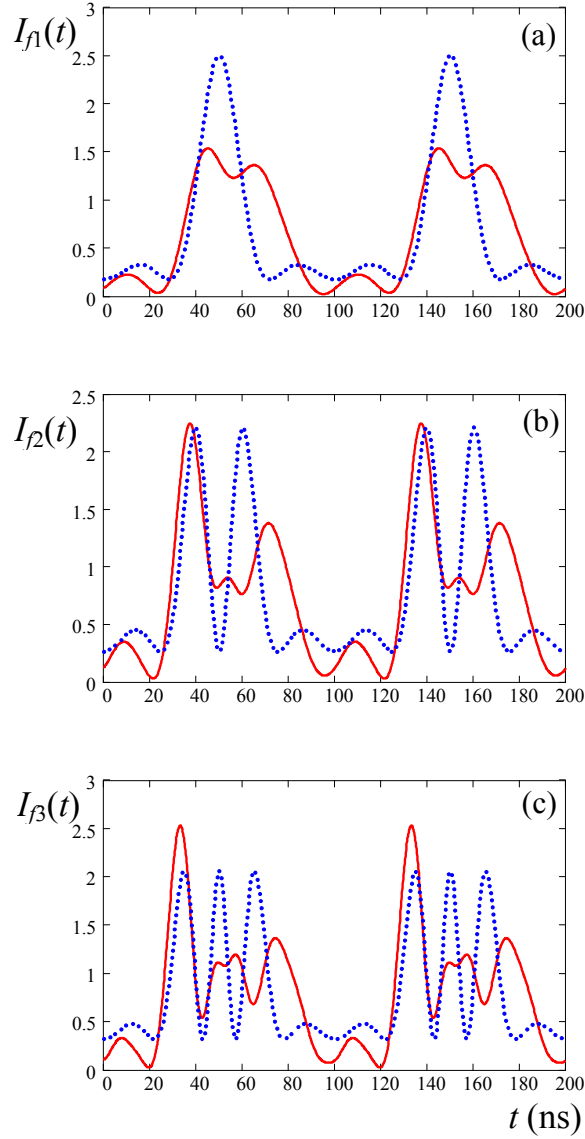}}\caption{Time
dependence of intensity of the radiation field filtered through the atomic
vapor with the optical depth $\alpha_{2}L=9053$ (solid line in red). Other
parameters and notations are the same as in Fig. 6}%
\label{fig:7}%
\end{figure}\begin{figure}[ptbptbptb]
\resizebox{0.5\textwidth}{!}{\includegraphics{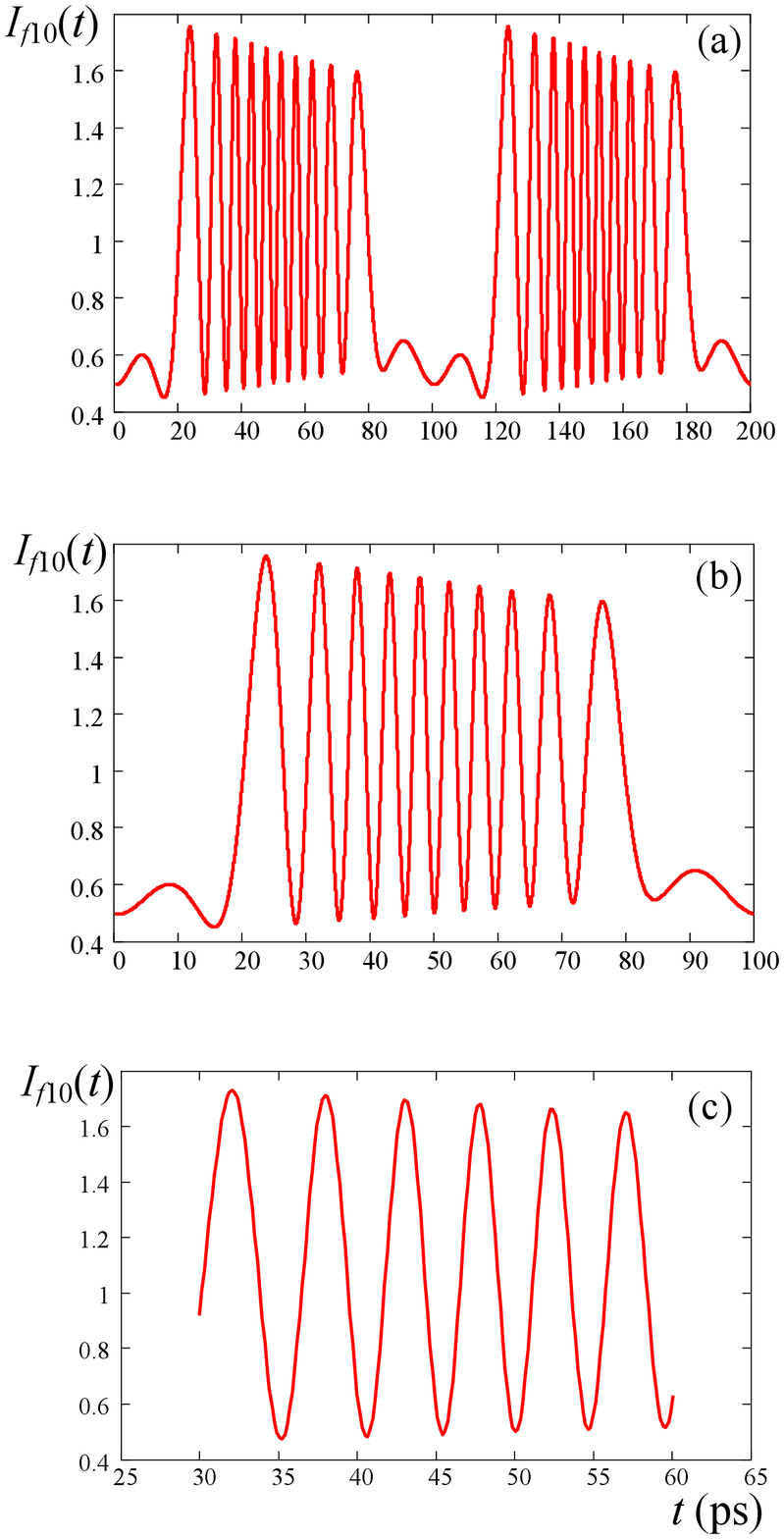}}\caption{(a) Time
dependence of intensity of the radiation field filtered through the atomic
vapor with the optical depth $\alpha_{0}L=453$. The spectral component $n=10$
is tuned in resonance with the filter and the phase modulation index is
$m_{10}=11.8$. (b) The content of a bunch of pulses. (c) Zoom in a central
part of the bunch. Other parameters and notations are the same as in Fig. 6}%
\label{fig:8}%
\end{figure}

For the Rb cell with atomic density $N_{1}=6\times10^{10}$ cm$^{-3}$ it is
enough to take into account the contribution of two spectral components
($n\pm1$) neighboring the resonant component $n$. For this density the
effective optical depth of the cell at the absorption line center is
$\alpha_{1}LF_{D}(0)=14.4$. The result of the comb filtering through the cell
is shown in Fig. 6. If $n=1$, $2$, or $3$ spectral component is in resonance
with the filter, the pulses with the width 25, 12.5, and 8.3 ps, respectively,
are produced.

If atomic density is increased by the order of magnitude to $N_{2}%
=6\times10^{11}$ cm$^{-3}$, then eight sidebands of the resonant component
$n\pm1$, $n\pm2$, $n\pm3$, and $n\pm4$ give noticeable contribution.
Therefore, the intensity modulation of the filtered comb becomes messy (see
Fig. 7). Actually, for $n=2$ and $n=3$ it is necessary to take into account
the contribution of 14 sidebands (up to $n\pm7$).

To obtain nice and clean pulses it is preferable to use a filter with a
smaller optical depth. An example of filtering by a cell with optical depth
$\alpha L=453$ the spectral component $n=10$ of the frequency comb, produced
by phase modulation with the optimal value of the modulation index
$m_{10}=11.8$ (which is close to $2\pi$), is shown in Fig. 8. Such a filtering
is capable to produce pulses as short as $2.5$ ps, which are grouped in
bunches consisting of 10 pulses. Thus, by phase modulation technique and
subsequent filtering it is possible to create pulses whose duration is 40
times shorter than the modulation period.

\section{Femtosecond pulses}

It is also possible to create femtosecond pulses if high order harmonic of the
frequency comb is removed. For simplicity we consider an ensemble of two-level
atoms, for example, a vapor of alkaline atoms. We take the modulation
frequency of the radiation field equal to $\Omega/2\pi=10$ GHz. If the
modulation index is $m_{100}=104$, the amplitude of the frequency component
$\omega_{r}-100\Omega$ takes its first maximum value, which is proportional to
$J_{100}(104)=0.144$. Removal of this spectral component of the comb by atomic
filter leads to high frequency oscillations of the transmitted field
intensity. Duration of the pulses is estimated as $T_{EO}/400=250$ fs, where
$T_{EO}$ is the phase modulation period. The necessary modulation index
$m_{100}=104$ corresponds to a voltage, which is $33$ times larger than the
have-wave voltage of electro-optical modulator. One can expect to reach this
value by increasing the voltage and/or the physical length of the modulator by
an order of magnitude.

Below the simplified picture of the filtering of high order harmonics of the
frequency comb is illustrated. To avoid complicated expressions, we
approximate the coherently scattered component of the field in Eq. (\ref{Eq6})
as proportional to $S_{100}=J_{100}(m_{100})$, assuming that the resonant
absorption is close to $100\%$, i.e., $T_{D}(0)\approx0$. We disregard the
dispersive contribution of the neighboring spectral components. Time
dependence of the intensity of the filtered radiation field is shown in Fig. 9
(a and c). The contrast of the pulses is not as large as for filtering of low
order harmonics. This is because the intensity of the maxima and minima of the
pulses can be estimated as $I_{\max}\approx\lbrack1+2J_{100}(104)]I_{0}$ and
$I_{\min}\approx\lbrack1-2J_{100}(104)]I_{0}$. Since $J_{100}(104)=0.144$, the
maxima exceeds $30\%$ the level of the intensity of the incident radiation
field, while minima are smaller than $I_{0}$ on $30\%$ of its value.

The amplitudes of high harmonics of the frequency comb decrease with increase
of their number $n$ since the first maxima of the Bessel functions
$J_{n}(m_{n})$ decrease with increase of the order $n.$ To achieve large
contrast of the pulses, one can remove several spectral components of the comb
neighboring $n=100$, if they have the same phase at time of the pulse
formation. As an example we consider the case if filtering of the comb
component $n=100$ is accompanied by filtering of the components with the
numbers $n+2k$, where $k$ takes values $\pm1$ and $\pm2$. These spectral
components have the same phase as $n=100$ spectral component at times when
central pulses of the bunch are formed. The amplitude of the filtered
radiation filed with cumulative contribution of extra four spectral components
of the comb due to their removal by separate filters is approximated as
\begin{equation}
E_{cum}(t)=E(t)\left[  e^{im\sin\Omega}-\sum_{k=-2}^{2}J_{n+2k}%
(m)e^{i(n+2k)\Omega t}\right]  , \label{Eq22}%
\end{equation}
where, for simplicity, it is supposed that $T_{D}(0)=0$. The intensity of the
filtered radiation field $I_{cum}(t)=\left\vert E_{cum}(t)\right\vert ^{2}$ is
shown in Fig. 9 (b and c). The central part of the pulse bunches demonstrates
good contrast. The maximum accumulative effect of five spectral components is
achieved if $m=103$. \begin{figure}[ptb]
\resizebox{1\textwidth}{!}{\includegraphics{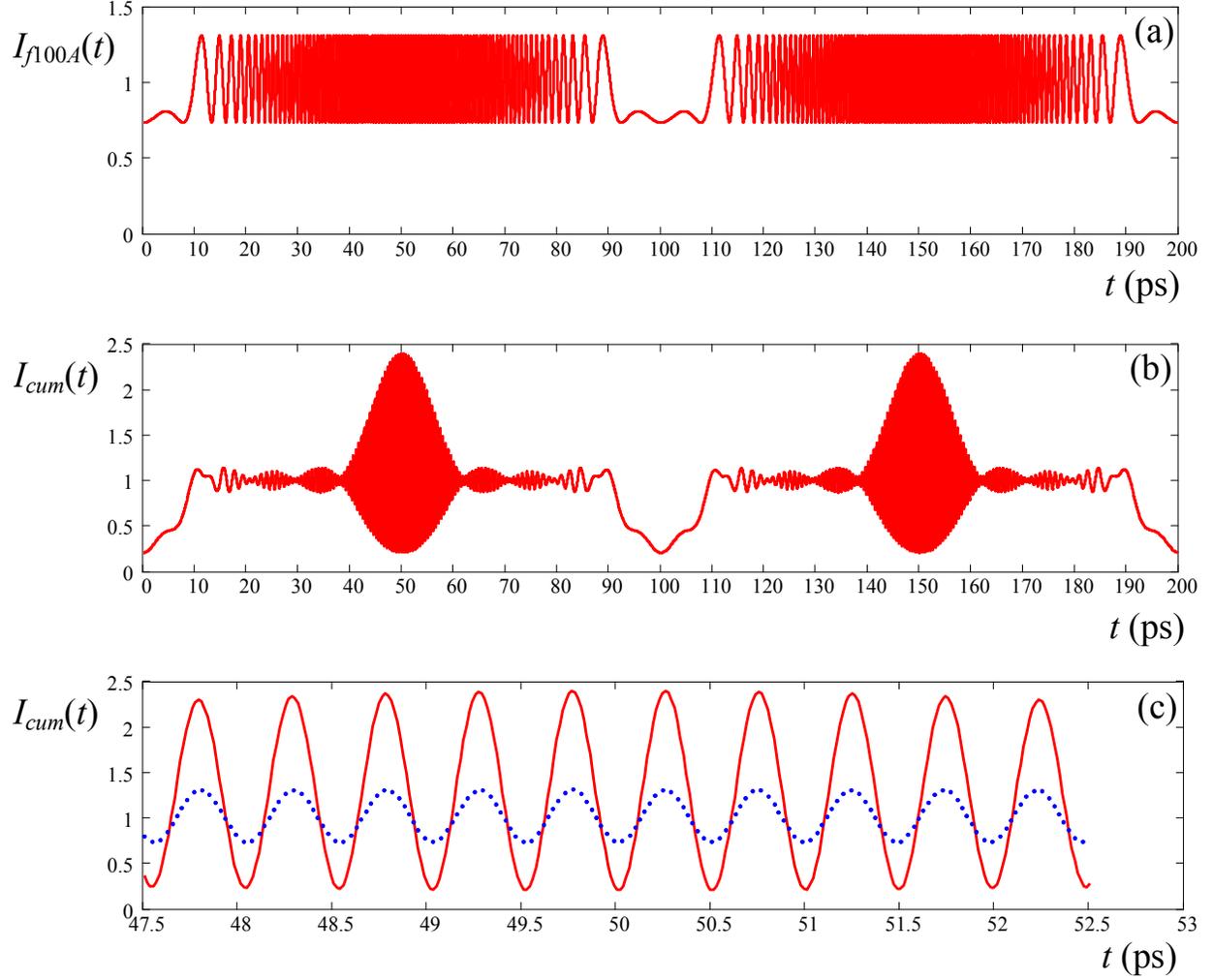}}\caption{(a) Time
dependence of the intensity of the radiation field whose $n=100$ spectral
component is removed by the filter. The modulation frequency is $\Omega
/2\pi=10$ GHz. (b) Accumulative result of the filtering of the additional four
spectral components of the frequency comb (see the text for details). (c) Zoom
in a central part of the bunch. Accumulative effect is shown by solid line (in
read), while dotted line (in blue) demonstrates the result of the filtering of
only one spectral component with $n=100$. Full width at half-maximum of the
pulses is close to $250$ fs as it is expected from simple estimations.}%
\label{fig:9}%
\end{figure}

Cumulative filtering is possible if we have narrow bandwidth filters properly
adjusted for chosen spectral components of the frequency comb. By a set of
cells with atomic vapors it is hard to construct such a specific
multifrequency filter having a large spacing between the absorptive frequency
components of the order of $20$ GHz. However, for example, cesium atoms have a
large hyperfine splitting of the ground state seen as a spectral doublet with
a spacing equal to $9.2$ GHz. If we reduce the modulation frequency of the
field phase down to $\Omega/2\pi=4.6$ GHz, populate properly ground state
levels by pumping Cs atoms such that both components of the doublet are
strongly absorptive, then we could remove two spectral components of the
frequency comb. An example of cumulative filtering of two spectral components
$100\Omega$ and $98\Omega$ by cesium atoms is shown in Fig. 10 (b and c). The
increase of the intensity of the pulses is clearly seen. Since the modulation
frequency is decreased by a factor of two, the duration of pulses increases by
the same factor to $500$ fs.

Cumulative filtering of two spectral components by atoms with the spectral
doublet produces pulses with modest contrast. The ratio of the maxima of the
pulse intensities to their minima is close to three. To make pulse minima
close to zero one can use destructive interference of the filtered radiation
field $E_{cum}(t)$ with the field $E_{EO}(t)$ from the same electro-optic
modulator but with the opposite phase and reduced amplitude, i.e.,%
\begin{equation}
E_{rdc}(t)=E_{cum}(t)-RE_{EO}(t), \label{Eq23}%
\end{equation}
where $R$ is the amplitude reduction factor. For the case of the filtering
with the doublet the reduction factor is $R=1-J_{100}(103)-J_{98}(103)$. Then,
due the interference the pulse minima at the bunch center become zero, while
the amplitude of the pulse maxima reduces to the value $2[J_{100}%
(103)-J_{98}(103)]E_{0}=0.55E_{0}$. The result of this interference is shown
in Fig. 10 (d and e) for the field intensity $I_{rdc}=\left\vert
E_{rdc}(t)\right\vert ^{2}$. \begin{figure}[ptb]
\resizebox{0.65\textwidth}{!}{\includegraphics{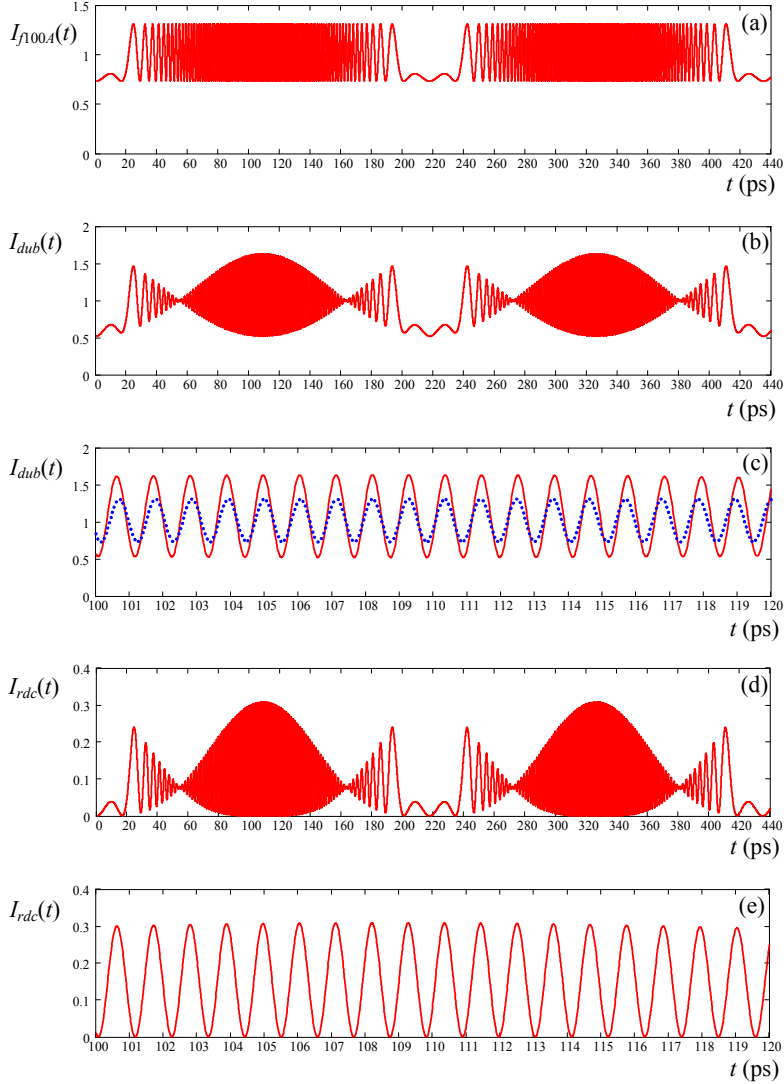}}\caption{(a) Time
dependence of the intensity of the radiation field whose $n=100$ spectral
component is removed by the filter. The modulation frequency is $\Omega
/2\pi=4.6$ GHz and modulation index is $m_{100}=104$. (b) Accumulative result
of the filtering of the additional spectral component $n=98$ of the frequency
comb by the doublet (see the text for details). The modulation index is
$m=103$. (c) Zoom in a central part of the bunch. Accumulative effect is shown
by solid line (in read), while dotted line (in blue) demonstrates the result
of the filtering of only one spectral component with $n=100$. Full width at
half-maximum of the pulses is close to $500$ fs as it expected from simple
estimations. (d) The result of the filtering with the doublet, followed by
reduction of the field amplitude due to the destructive interference with the
reduced field from electro-optic modulator whose phase is shifted by $\pi$.
(e) Zoom in the central part of the pulse bunch, shown in (d).}%
\label{fig:10}%
\end{figure}

Cumulative filtering is better to implement by organic molecules doped in
polymer matrix, which undergo persistent spectral hole burning at liquid
helium temperature. In such a filter the frequency resolution is limited by
the width of the homogeneous zero-phonon lines of the chromophore molecules.
For example, waveguide narrowband optical filter, which consists a planar
waveguide with a thin polymer film containing molecules, which undergo
persistent spectral hole burning at liquid helium temperature, demonstrates
transmission bandwidth less than 1 GHz \cite{Tschanz1995,Tschanz1996}.
Saturated holes are burned in waveguide geometry by illumination in the
transverse direction with low absorption, whereas the probing is carried out
in longitudinal wave guiding directions with high absorption. The waveguide
with spectral hole burning can act as integrated sub-gigahertz narrow-band
filter, which is proposed to observe slow light phenomenon
\cite{Shakhmuratov2005,Rebane2007}.

Comb structures of arbitrary shapes in transmission spectra were created
experimentally in organic molecules doped in a polymer
\cite{Rebane1995,Wild2002}. Therefore, one can expect that it is
experimentally possible to create a broad hole with, say, five absorptive
peaks in it. For cumulative filtering it is enough to create a broad hole with
the width $\sim10$ cm$^{-1}$ and five absorptive peaks in it with $20$ GHz
spacing and widths less than $1$ GHz.

\section{Coherent Raman scattering with resonant filtering}

Frequency modulation (FM) of the field, followed by a dispersive compensator
to produce short pulses, could be substituted by dispersive modulators
\cite{Loy1,Loy2,Grischkowsky1975}. In dispersive modulators, where instead of
modulating the frequency of the light, the resonant frequency of the filter is
modulated. Both techniques, FM followed by dispersive compensator and
dispersive modulator, use light-matter interaction, which results in a phase
change of the field spectral components, being far from resonance. Here I
propose to make a specific resonant filtering in the interaction process of
light with particles whose resonant frequencies experience modulation.
Theoretical description of this process is very similar to that, proposed in
\cite{Grischkowsky1975}. The only difference is the resonant condition imposed
on one spectral component.

In the linear response approximation the propagation of the CW radiation field
$E(z,t)=E_{0}\exp(-i\omega_{r}t+ikz)$ through the resonant medium is described
by the set of equations%
\begin{equation}
\widehat{L}\chi(z,t)=i\alpha\gamma\sigma_{eg}(z,t)/2, \label{Eq24}%
\end{equation}%
\begin{equation}
\dot{\sigma}_{eg}=(i\Delta-\gamma)\sigma_{eg}+i\chi(z,t), \label{Eq25}%
\end{equation}
where $\widehat{L}=\partial_{z}+c^{-1}d_{t}$, $\chi(z,t)=d_{eg}E_{0}%
(z,t)/2\hbar$ is the Rabi frequency of the field proportional to the slowly
varying field amplitude $E_{0}(z,t)=E(z,t)\exp(i\omega_{r}t-ikz)$, which
satisfies the boundary condition $E_{0}(0,t)=E_{0}$ at the entrance to the
medium with coordinate $z=0$; $d_{eg}$ is a matrix element of the dipole
transition between ground, $g$, and excited, $e$, states of a molecule
(NH$_{3}$ in the experiment of Loy \cite{Loy1}); $\rho_{eg}=\sigma_{eg}%
\exp(-i\omega_{r}t+ikz)$ is the nondiagonal element of the molecular density
matrix and $\sigma_{eg}$ its slowly varying part; $\gamma$ is the decay rate
of the coherence of the states $e$ and $g$, $\Delta=\omega_{r}-\omega_{0}$ is
the detuning from the resonant frequency $\omega_{0}$ of molecules, and
$\alpha$ is the resonant absorption coefficient of the molecular vapor.

The resonant frequency of molecules is externally controlled via the Stark
effect. Long Stark sell, with a small plate separation, is a parallel-plate
transmission line. A bias electric filed is used to control the dc frequency
separation between the input light and the relevant molecular transition. An
applied rf voltage travels in the same direction as the light. Therefore, the
resonant detuning is $\Delta=\Delta_{b}+\Delta_{rf}\cos\Omega t$, where
$\Delta_{b}$ is the dc offset between the radiation frequency and the
absorption line center due to the bias voltage, $\Omega$ is the rf voltage
frequency, and $\Delta_{rf}$ is the amplitude of the time-dependent frequency modulation.

For this parallel plate transmission line the Stark voltage travels at $c$ to
a very good approximation. Therefore spatiotemporal dependence of $\Delta$
satisfies the equation%
\begin{equation}
\widehat{L}\Delta(z,t)=0, \label{Eq26}%
\end{equation}
whose solution is $\Delta(z,t)=\Delta(t-z/c)$.

With the help of new variables%
\begin{equation}
\chi_{comb}(z,t)=\chi(z,t)e^{im\sin\Omega t^{\prime}}, \label{Eq27}%
\end{equation}%
\begin{equation}
p_{eg}(z,t)=\sigma_{eg}(z,t)e^{im\sin\Omega t^{\prime}}, \label{Eq28}%
\end{equation}
equations (\ref{Eq24}) and (\ref{Eq25}) are reduced to
\begin{equation}
\widehat{L}\chi_{comb}(z,t)=i\alpha\gamma p_{eg}(z,t)/2, \label{Eq29}%
\end{equation}%
\begin{equation}
\dot{p}_{eg}=(i\Delta_{b}-\gamma)p_{eg}+i\chi_{comb}(z,t), \label{Eq30}%
\end{equation}
where $t^{\prime}=t-z/c$ is the local time and $m=\Delta_{rf}/\Omega$.

There are no time dependent coefficients in these equation. Therefore, their
steady state solution can be found in the form%
\begin{equation}
\chi_{comb}(z,t)=%
%TCIMACRO{\dsum \limits_{n=-\infty}^{+\infty}}%
%BeginExpansion
{\displaystyle\sum\limits_{n=-\infty}^{+\infty}}
%EndExpansion
\chi_{n}(z)e^{in\Omega t^{\prime}}, \label{Eq31}%
\end{equation}%
\begin{equation}
p_{eg}(z,t)=%
%TCIMACRO{\dsum \limits_{n=-\infty}^{+\infty}}%
%BeginExpansion
{\displaystyle\sum\limits_{n=-\infty}^{+\infty}}
%EndExpansion
p_{n}(z)e^{in\Omega t^{\prime}}, \label{Eq32}%
\end{equation}
where%
\begin{equation}
p_{n}(z)=\frac{-\chi_{n}(z)}{\Delta_{b}-n\Omega+i\gamma}, \label{Eq33}%
\end{equation}%
\begin{equation}
\chi_{n}(z)=T_{n}(z)\chi_{n}(0), \label{Eq34}%
\end{equation}%
\begin{equation}
T_{n}(z)=\exp\left[  \frac{-\alpha z/2}{1-i(\Delta_{b}-n\Omega)/\gamma
}\right]  , \label{Eq35}%
\end{equation}
$\chi_{n}(0)=\chi(0)J_{n}(m)$, and $\chi(0)=d_{eg}E_{0}/2\hbar$.

For a Doppler-broadened line, whose homogeneous linewidth $2\gamma$ is much
smaller than the Doppler width $\Delta\omega_{D}$, the transmission function
$T_{n}(z)$ is modified as in Sec. IV [see Eqs. (\ref{Eq15}), (\ref{Eq16})].

In Loy's experiments Doppler width $\Delta\omega_{D}/2\pi=87$ MHz was larger
than rf modulation frequency $\Omega$, which varied between $30$ and $50$ MHz.
The dc offset $\Delta_{b}$ was comparable or much larger than $\Delta
\omega_{D}$. Therefore, the spectral components of the frequency comb
$\chi_{comb}(z,t)$ mainly acquired different phase shifts at the exit of the
dispersive filter. This is the core idea of the dispersive modulators.

We propose to employ modulation frequency, which is much larger than Doppler
width (for example, $\Omega/2\pi=300$ MHz, which is almost an order of
magnitude larger than $\Delta\omega_{D}/2$). Then, if one of the spectral
components of the comb $\chi_{comb}(z,t)$ comes to resonance with molecular
vapor, pulse bunches are developed at the exit of the modulating filter
similar to those shown in Fig. 4.

The physical origin of the pulse formation in our case can be explained as
follows. If all spectral components of the comb are far from resonance with
the vapor, one can neglect their change assuming that $\chi_{n}(z)\approx
\chi_{n}(0)$. Then, we have%
\begin{equation}
\chi_{comb}(z,t)=\chi(0)%
%TCIMACRO{\dsum \limits_{n=-\infty}^{+\infty}}%
%BeginExpansion
{\displaystyle\sum\limits_{n=-\infty}^{+\infty}}
%EndExpansion
J_{n}(m)e^{in\Omega t^{\prime}}=\chi(0)e^{im\sin\Omega t^{\prime}}.
\label{Eq36}%
\end{equation}
Applying transformation inverse to (\ref{Eq27}), we obtain no change of the
radiation field at the exit of the filter, i.e., $\chi(z,t)=\chi(0,t^{\prime
})$.

If $n$-th spectral component of the comb is removed by the filter, then the
comb changes as%
\begin{equation}
\chi_{comb}(z,t)=\chi(0)\left[  e^{im\sin\Omega t^{\prime}}-J_{n}%
(m)e^{in\Omega t^{\prime}}\right]  . \label{Eq37}%
\end{equation}
In the laboratory reference frame (after the inverse transformation) we obtain%
\begin{equation}
\chi(z,t)=\chi(0)\left[  1-J_{n}(m)e^{in\Omega t^{\prime}-im\sin\Omega
t^{\prime}}\right]  . \label{Eq38}%
\end{equation}
Constructive interference of the time dependent component with the constant
one results in formation of pulses, while their destructive interference gives
dark windows. This equation can be interpreted as a resonant interaction of
the $n$-th component with molecules followed by Raman rescattering in all the
sidebands. In this picture the primary source of the radiation field is the
polarization excited by resonant component $\chi(0)J_{n}(m)\exp(in\Omega
t^{\prime})$. This excitation is rescattered coherently into $\chi
(0)J_{n}(m)J_{k}(m)\exp[i(n-k)\Omega t^{\prime}]$ components (with $k$ varied
from $+\infty$ to $-\infty$) due to the modulation of the resonant frequency
of the molecules.

\section{Discussion}

FM followed by dispersive compensator scheme for producing optical pulses have
been based on the ideas of chirp radar
\cite{Klauder1960,Duguay1966,Duguay1968}. In the chirp radar system the
transmitted pulse is of relatively long duration $\Delta t$ during which time
the instantaneous frequency is swept over the range $\omega\rightarrow
\omega+\Delta_{chirp}$ satisfying the condition $\Delta_{chirp}\Delta t\gg1$.
The return pulse is passed through a dispersive network providing a
differential delay $\Delta t$ over the frequency range $\Delta_{chirp}$. As a
result, the energy at the beginning of the pulse is delayed so as to reach the
end of the network at the same time as the energy at the end of the pulse. The
duration of the compressed pulse produced this way is of the order of $\delta
t\approx1/\Delta_{chirp}\ll\Delta t$.

In optical domain electrooptical modulation of the radiation frequency
\cite{Grischkowsky1974} spreads its spectrum over the range $m\Omega$. In
Loy's dispersive filter rf voltage due to Stark effect spreads effectively the
field spectrum over the range $\Delta_{rf}=m\Omega$. In both schemes
dispersive properties of atoms or molecules are used to have
frequency-dispersive group velocities of the spectral components produced by
modulation. Therefore, both schemes require resonant media with a very large
optical thickness and employ large offset between the radiation frequency and
the absorption line center to avoid the absorptive losses. These constrains
force to work with relatively small frequencies $\Omega$ and large modulation
index $m$. Then, many spectral components of the comb acquire appreciable
dispersive phase shifts with large difference between blue and red borders of
the comb spectrum.

It is quite easy to estimate parameters of the dispersive filter, which is
capable to compress phase modulated field into short pulses within the chirp
radar model. We take as an example the parameters used in the experiment of
Pearson \textit{et al.} \cite{Pearson} who compressed phase modulated CW
radiation field by passing it trough a sodium vapor. Resonant detuning
$\Delta_{c}/2\pi=(\omega_{r}-\omega_{f})/2\pi=4$ GHz was large compared with
the phase modulation frequency $\Omega/2\pi=200$ MHz. The modulation index
$m\approx2\pi$ was moderate.

If the field phase is modulated according to $\varphi(t)=m\sin\Omega t$, then
the effective resonant detuning evolves in time as $\Delta(t)=\Delta
_{c}-m\Omega\cos\Omega t$. Half a period this detuning increases taking
maximum value $\Delta_{\max}/2\pi=(\Delta_{c}+m\Omega)/2\pi=5.26$ GHz at
$t_{\max}=(2n+1)\pi/\Omega$, another half it decreases taking minimum value
$\Delta_{\min}/2\pi=(\Delta_{c}-m\Omega)/2\pi=2.74$ GHz at $t_{\min}%
=(2n+2)\pi/\Omega$, where $n$ is an arbitrary integer number. Portion of
radiation field with frequencies detuned from resonance by $\Delta_{\max}$
travels through the vapor cell with group velocity larger than that portion,
which is detuned by $\Delta_{\min}$. If the latter is produced at earlier
time, for example, at $t_{\min}=2n\pi/\Omega$, and the former at later time,
for example, at $t_{\max}=(2n+1)\pi/\Omega$, they have a chance to arrive at
the exit of the cell at the same time if the difference of their group
velocities is appropriate. The same is true for all the intermediate spectral
components of the field. Such a compression of spectral components of the
field happens between $t_{\min}=2n\pi/\Omega$ and $t_{\max}=(2n+1)\pi/\Omega$.
In the next half a period of phase modulation, between $t_{\max}%
=(2n+1)\pi/\Omega$ and $t_{\min}=(2n+2)\pi/\Omega$, these spectral components
are spread since the late component comes much later than the first component.
In this half a period the difference of the group velocities of the spectral
components of the field results in a drop of the radiation intensity.

Delay time of the spectral components can be estimated as $\tau_{d}=\alpha
L\gamma/2\Delta^{2}$, where $\Delta$ is the resonant detuning of the spectral
component \cite{Shakhmuratov2008}. Then, we have for two spectral components
$\Delta_{\min}$ and $\Delta_{\max}$ two delay times $\tau_{d\min}$ and
$\tau_{d\max}$. To observe the pulse compression we need to satisfy the
condition $\tau_{d\min}-\tau_{d\max}=T_{EO}/2$, where $T_{EO}=2\pi/\Omega$ is
a phase modulation period. In the experiment \cite{Pearson} this condition is
$\tau_{d\min}-\tau_{d\max}=2.5$ ns. It is realized if $\alpha L=6.5\times
10^{4}$.

To verify these estimations we use simplified expression for the filtered
field amplitude%
\begin{equation}
E_{f}(t)=E(t)%
%TCIMACRO{\dsum \limits_{n=-\infty}^{+\infty}}%
%BeginExpansion
{\displaystyle\sum\limits_{n=-\infty}^{+\infty}}
%EndExpansion
J_{n}(m)e^{in\Omega t-\frac{\alpha L\gamma/2}{\gamma-i(\Delta_{c}-n\Omega)}},
\label{Eq39}%
\end{equation}
where $\gamma$ is the homogeneous dephasing rate ($\gamma/2\pi=5$ MHz). Here
we disregard the Doppler broadening because of large detuning $\Delta_{c}$
(see the arguments in Sec. IV).

Time dependence of the intensity of the filtered radiation field
$I_{f}(t)=\left\vert E_{f}(t)\right\vert ^{2}$, which is described by Eq.
(\ref{Eq39}), is shown in Fig. 11 (solid lines in red) for the parameters
specified above. The pulses look quite similar to those shown in Fig. 2(b) in
Ref. \cite{Pearson}. Time dependence of the intensity for $\alpha L$ two times
smaller than the optimal value, Fig. 11(a), and two times larger, Fig. 11 (b)
are shown by dotted lines (in blue) for comparison.

It is instructive to express the field amplitude $E_{f}(t)$, taking into
account finite number of terms of the absorption line Taylor series
\begin{equation}
\frac{\alpha L\gamma/2}{\gamma-i(\Delta_{c}-n\Omega)}\approx i\frac{\alpha
L\gamma}{2\Delta_{c}}\left(  1+\frac{n\Omega}{\Delta_{c}}+\frac{n^{2}%
\Omega^{2}}{\Delta_{c}^{2}}+\frac{n^{3}\Omega^{3}}{\Delta_{c}^{3}}+...\right)
, \label{Eq40}%
\end{equation}
where the contribution of the coherence decay rate $\gamma$ in the denominator
is disregarded since $\gamma\ll\Delta_{c}$. The first term of this series
describes the constant phase shift of the field. The second term reduces the
group velocity of the pulsed field. The third and fourth terms result in
dispersion and higher order dispersion of the group velocities of the spectral
components of the pulse.

If we take into account only three terms of the series, then for the specified
values of the parameters Eq. (\ref{Eq39}) is reduced to%
\begin{equation}
E_{d}(t)=E(t)e^{i\varphi_{d}}%
%TCIMACRO{\dsum \limits_{n=-\infty}^{+\infty}}%
%BeginExpansion
{\displaystyle\sum\limits_{n=-\infty}^{+\infty}}
%EndExpansion
J_{n}(m)e^{in\left(  \Omega t-a-\varepsilon_{1}n\right)  }, \label{Eq41}%
\end{equation}
where $\varphi_{d}=-\alpha L\gamma/2\Delta_{c}$, $a=2.03$, and $\varepsilon
_{1}=0.102$. Time dependence of the intensity $I_{d}(t)=\left\vert
E_{d}(t)\right\vert $, where only first three terms of the series are taken
into account including dispersion of the group velocities, is shown in Fig.
11(c) by dotted line (in blue). Group velocity dispersion of the spectral
components of the phase modulated field in a dispersive filter results in a
compression of the field into a single pulse of short duration with two small
side lobes.

If we take into account also the contribution of the fourth term in the series
(\ref{Eq40}), which produces dispersion of the group velocity dispersion, then
Eq. (\ref{Eq39}) is reduced to%
\begin{equation}
E_{dd}(t)=E(t)e^{i\varphi_{d}}%
%TCIMACRO{\dsum \limits_{n=-\infty}^{+\infty}}%
%BeginExpansion
{\displaystyle\sum\limits_{n=-\infty}^{+\infty}}
%EndExpansion
J_{n}(m)e^{in\left(  \Omega t-a-\varepsilon_{1}n-\varepsilon_{2}n^{2}\right)
}, \label{Eq42}%
\end{equation}
where $\varepsilon_{2}=0.005$. Time dependence of the intensity $I_{dd}%
(t)=\left\vert E_{dd}(t)\right\vert $ is shown in Fig. 11(d) by dotted line
(in blue). The forth term results in a pulse ringing and reduction of the
pulse intensity. \begin{figure}[ptb]
\resizebox{0.7\textwidth}{!}{\includegraphics{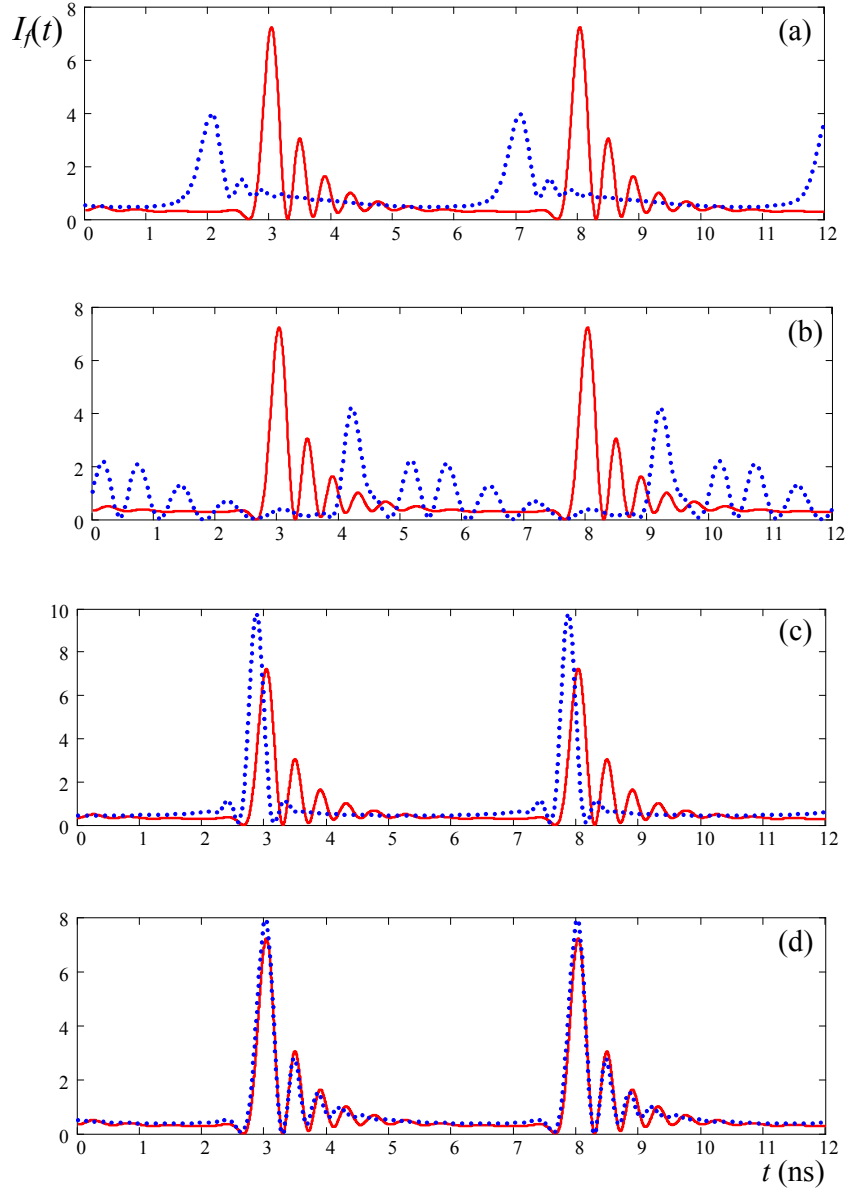}}\caption{Time
dependence of the intensity of the radiation field filtered through a
dispersive filter with different group velocities of the spectral components
of the phase modulated field, solid line (in red). Dotted line (in blue) shows
the same dependence but for the filter, which is two times shorter (a), longer
(b). Time dependence of the intensity (dotted line in blue) if we disregard
(c) or take into account (d) the forth term of Taylor series of the absorption
line, which is responsible for higher order dispersion of the group velocity
of spectral packets.}%
\label{fig:11}%
\end{figure}

According to the numerical analysis the first pulse, shown by dotted line (in
blue) in Fig. 11(c), takes its maximum value at $t_{p}$ when $\Omega
t_{p}-a=\pi/2$. Intuitively, this is consistent with the concept of the chirp
radar model. Time delay $t_{c}$ of the fundamental frequency component
$\omega_{r}$ ($n=0$) at the exit of the dispersive filter satisfies the
condition $\Omega t_{c}-a=0$ since the parameter $a$ is calculated for the
detuning $\Delta_{c}$, while the most detuned spectral component $\Delta
_{c}+m\Omega$ appears quarter of the period $T_{EO}$ later. Therefore, pulse
compression takes place when $\Omega t_{p}-a=\pi/2$. Individual phase shifts
of the spectral components $\omega_{r}-n\Omega$ with $n\neq0$ due to
dispersion of their group velocities are described by the term $in^{2}%
\varepsilon_{1}$ in the exponent in Eq. (\ref{Eq41}).

At time $t_{p}$ the field amplitude takes the value%
\begin{equation}
E_{d}(t_{p})=E(t_{p})e^{i\varphi_{d}}%
%TCIMACRO{\dsum \limits_{n=-\infty}^{+\infty}}%
%BeginExpansion
{\displaystyle\sum\limits_{n=-\infty}^{+\infty}}
%EndExpansion
i^{n}J_{n}(m)e^{-i\varepsilon_{1}n^{2}}, \label{Eq43}%
\end{equation}
where $i^{n}=\exp(in\pi/2)$. Numerical calculation of the sum in Eq.
(\ref{Eq43}) gives%
\begin{equation}%
%TCIMACRO{\dsum \limits_{n=-\infty}^{+\infty}}%
%BeginExpansion
{\displaystyle\sum\limits_{n=-\infty}^{+\infty}}
%EndExpansion
i^{n}J_{n}(m)e^{-i\varepsilon_{1}n^{2}}=2.22-2.26i, \label{Eq44}%
\end{equation}
and $I_{d}(t_{p})=9.62I_{0}$.

In the scheme of resonant filtering, proposed in the presented paper, only one
spectral component of the comb is removed by a resonant absorber. Optical
thickness of the absorber is to be modest to avoid phase change of nonresonant
components of the comb. Then, short pulses are produced. Their duration is
close to $\pi/2n\Omega$, which qualitatively differs from the pulse duration in
the dispersive compensator scheme, estimated according to the chirp radar as
$1/2m\Omega$. Quantitatively these durations are also different since the
optimal values of the modulation index in our scheme can be roughly estimated
as $m\approx\pi n/2$ for $n=\pm1,\pm2,\pm3$ and $m\approx n$ for large $n$. A
specific qualitative difference of the schemes is that the modulation
frequency in our scheme is to be large, at least much larger than the width of
the absorption line of the resonant filter. The modulation index $m$ is also
must be large to make pulses as short as possible. However, if the modulation
index exceeds $20$, several spectral components should be removed to keep
large contrast of the produced pulses. These components have to be spaced
apart from the target component $\omega-n\Omega$ by $\pm2\Omega$, i.e. the
accompanying spectral components to be removed are $\omega-(n\pm2)\Omega$ and
$\omega-(n\pm4)\Omega$. The filtering must be selective leaving untouched the
intermediate spectral components $\omega-(n\pm1)\Omega$ and $\omega
-(n\pm3)\Omega$.

\section{Conclusion}

The resonant method of converting phase modulation into amplitude modulation
of the radiation field in optical domain is discussed. Phase modulation could
be implemented by electro-optic modulator, which converts a single line
radiation field into a frequency comb consisting of fundamental frequency
$\omega_{r}$ and sidebands spaced apart at distances that are multiples of the
modulation frequency, i.e., $\omega_{r}\pm n\Omega$ ($n=1$, $2$, ...). The
intensity of the spectral components of the comb is proportional to $J_{n}%
^{2}(m)$, where $m$ is the phase modulation index. If the $n$-th spectral
component of the comb is tuned in resonance with atoms in the filter whose
resonant absorption line is narrower than the frequency spacing $\Omega$ of
the comb, then the filtered field is transformed into pulses demonstrating a
conversion of the phase modulation into intensity modulation. The effect is
explained by the interference of the coherently scattered resonant component
in the forward direction with the whole comb. Constructive interference of the
fields results in formation of pulses. Their destructive interference is seen
as dark windows. Four examples of the resonant filter are analyzed. They are
an ensemble of cold atoms, atomic vapor of alkaline atoms, NH$_{3}$ molecules
experiencing Stark modulation of the resonant frequency, and organic molecules
doped in polymer matrix. It is shown that it is preferable to work with the
filters with moderate optical depth. If the optical depth is large then due to
dispersive wings of the filter many spectral components of the comb are
modified producing the pulse corruption. Creation of nanosecond,
subnanosecond, and femtosecond pulses is analyzed. The number of pulses and
their spacing is well controlled by the modulation frequency and modulation
index. The proposed method could be also applied for photon shaping. Single
photons of spectral widths ranging from several MHz up to 1 GHz could be
modified to encode the information into time-bin qubits.

\section{Acknowledgments}

This work was partially funded by the Program of Competitive Growth of Kazan
Federal University, funded by the Russian Government, and the RAS Presidium
Program ``Actual problems of low temperature physics.''

\end{document}